\documentclass[letterpaper]{article}
\usepackage{iccc}
\usepackage{graphicx}
\usepackage{subcaption}
\usepackage{natbib}
\usepackage{dirtytalk}

\usepackage{placeins}
\usepackage{float}

\usepackage{times}
\usepackage{helvet}
\usepackage{courier}
\title{Towards Movement Generation with Audio Features}

\author{Benedikte Wallace$^{1}$, Charles P. Martin$^{2}$, Jim Torresen$^1$ \and Kristian Nymoen$^{1}$ \\
\mbox{}\\
$^1$RITMO Center, University of Oslo\\ %
$^2$Australian National University\\
benediwa@ifi.uio.no} %

\begin{document} 

\maketitle

\restylefloat{figure}

\begin{abstract}
\begin{quote}
Sound and movement are closely coupled, particularly in dance.
Certain audio features have been found to affect the way we move to music.
Is this relationship between sound and movement something which can be modelled using machine learning? 
This work presents initial experiments wherein
high-level audio features calculated from a set of music pieces are included in a movement generation model trained on motion capture recordings of improvised dance. 
Our results indicate that the model learns to generate realistic dance movements which vary depending on the audio features.
\end{quote}
\end{abstract}

\section{Introduction}
Expressive movement is an intrinsic part of human life. 
Hand gestures, body language as well as dance can efficiently convey an emotional state. 
Simple movement patterns such as gait or arm movement may allow us to detect characteristics such as personality or mood \citep{michalak2009embodiment, satchell2017evidence, pollick2001perceiving}. 
As such, a better understanding of body motion, and the analysis and generation of motion data is important to further develop fields such as 
human-robot interaction and human activity recognition. For dance movement in particular, generative models have potential as artistic tools for animation and choreography.  

Research in embodied music cognition has identified several audio features that are relevant to how we move to music.
\citeauthor{Burger2013}'s \citeyearpar{Burger2013} work suggests that several mappings exist between different aspects of music and music-induced movement. 
The presence of a clear beat, for example, was shown to translate to faster movements of head and hands.

The work presented here is part of an ongoing research effort to examine how deep learning can be used to capture salient features of human movement, and especially dance movement, using full-body motion capture data and sound.
As part of this work, we have collected a dataset of motion capture recordings of dance improvisation performed to six different musical stimuli.

Here, we present the results of training a generative mixture density recurrent neural network (MDRNN) on 
our motion data and audio features which have been shown to affect certain aspects of movement to music.
Without the inclusion of audio features, the MDRNN is able to generate sequences of movement which are (subjectively) realistic variations of the underlying training data.
The results presented here indicate that the model retains this ability to produce movement variations when audio features are added. Our findings suggest that the model additionally learns that different audio features affect the way the body moves.

\begin{figure}
    \centering
    \includegraphics[scale=.20]{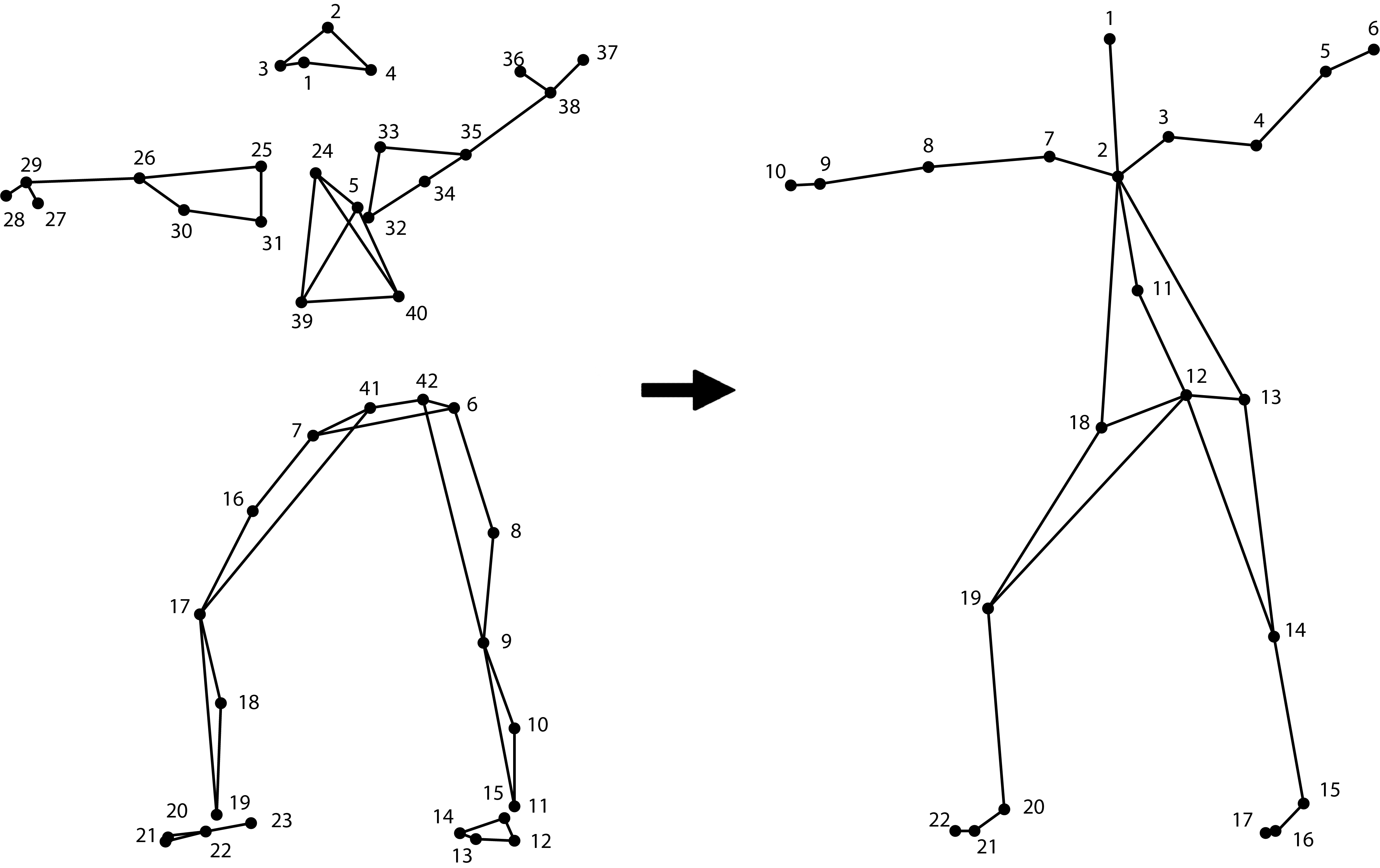}
    \caption{The 43 reflective markers translated to 22 points}
    \label{fig:joints}%
\end{figure}

\section{Motion Capture Data}
Our dataset contains 54 one minute motion capture recordings of improvised dance performed by three experienced dancers. 
Each dancer performs three one minute improvisations to six different musical stimuli which vary in terms of tempo, tonality and beat presence. 
The dataset was recorded using a Qualisys optical motion capture system with 12 Oqus 300/400 series cameras which capture 43 reflective markers worn by the dancers. 
Figure \ref{fig:joints} shows how the 43 marker positions were reduced to a 22 point skeleton representation using the MoCap Toolbox 1.5 \citep{burger2013mocap}. 
Small gaps in the data were spline-filled using Qualisys Track Manager 2019.3 and a 2nd degree Butterworth filter with a .03Hz cutoff was applied to remove any marker jitter.

Recordings in our dataset have been normalized so that the root marker (a weighted average of markers 41, 42, 6 and 7 in Figure \ref{fig:joints}) is centred at the origin.
Body segment lengths are averaged across the three dancers ensuring that the data is invariant to global position and individual body dimensions. 
The data was captured at 240Hz and downsampled to 30Hz before model training to reduce the size of each example. 
The resulting 54 data tensors consist of 1800 frames (60 seconds at 30Hz) with 3-dimensional positions for each of the 22 points.

Two full motion capture recordings were withheld for testing while the remaining 52 examples were split into two sets, 80\% were used for training and 10\% for validation. 
Each example has been sliced into overlapping sequences of 300 frames and the spatial dimensions of each of the 22 points are scaled using min-max normalisation.
The input to our model thereby consists of 78000 overlapping sequences of 300 frames with their corresponding audio features. 
The model performs a sequence-to-sequence mapping between the training examples (including audio features) and 
the shifted sequence of motion capture frames (excluding audio features -- the model only predicts motion).

\section{Representing sound}

There are several aspects to consider when selecting appropriate audio features to represent the music examples to which the dancers were improvising. 
Several previous works \citep{seo2013autonomous, fukayama2015music, lee2019dancing} have largely focused on the rhythmical, beat-matching aspects when generating dance.

Although the presence of a clear beat can affect our urge to move, moving in sync with a beat is only one of several ways musical features influence the way we move.
For the experiments presented here, we have chosen to use two high-level rhythm- and timbre-related features: pulse clarity and sub-band spectral flux.
Previous work by \citep{Burger2013} has shown connections between pulse clarity \citep{lartillot2008multi} and overall body movement, as well as sub-band spectral flux and movement of the head and hands.  

Pulse clarity is a high-level feature which measures how clearly the underlying pulse of the music is perceived. 
Pulse clarity is estimated using the overall entropy of the energy distribution of the frequency spectrum within a musical piece. 
We calculate a series of pulse clarity values for each musical stimuli using a sliding window of 5 seconds and a hop size of 0.08 seconds. This gives us a time series wherein each value corresponds to a single frame of the motion capture data.

Spectral sub-band flux measures spectral changes in different frequency bands of an audio signal. 
\citep{alluri2010exploring} found that the sub-band fluctuations in the region between 50 Hz and 200 Hz are related to the perceived \say{fullness} of a musical piece, while fluctuations in the region of 1600 Hz and 6400 Hz were linked to the perceived \say{activity} of the piece. 
These sub-bands also correspond to activity from rhythmic instruments such as kick drum and bass guitar for the lower frequency band and hi-hat and cymbals for the higher range.

We extract two frequency bands, one low-frequency band (50 Hz - 100 Hz) and one high-frequency band (3200 Hz - 6400 Hz) from the six musical stimuli.
The spectral flux is then calculated using the same window and hop size as for the pulse clarity values resulting in a single sub-band flux value for each of the two bands for every frame of motion capture data. 
The audio features are appended to each data frame of the motion capture data to create the tensors used to train the generative mixture density recurrent neural network.

\begin{figure}
    \centering
    \includegraphics[width=\linewidth]{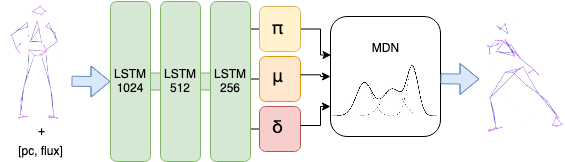}
    \caption{Sampling from the MDRNN. One frame of mo-cap data and audio features is sent through the model. The MDRNN outputs the parameters of a mixture distribution which is sampled to generate the next frame.}
    \label{fig:mdn}
\end{figure}

\section{Mixture Density Recurrent Neural Networks}
Mixture density networks (MDNs)~\citep{astonpr373} treat the outputs of a neural network as the parameters of a Gaussian mixture model (GMM), which can be sampled to generate real-valued predictions. 
A GMM can be derived using the mean, weight and standard deviation of each component. The number of components needed to accurately represent the data is not known and is treated as a hyperparameter for our model. 
For the study outlined here, we have used 3 components. 
By combining a recurrent neural network (RNN) with an MDN to form an MDRNN we can make real-valued predictions based on a sequence of inputs. 
Figure \ref{fig:mdn} shows the model architecture of the MDRNN used in this work. 
The RNN consists of three layers of LSTM cells \citep{Schmidhuberlstm}.
The three LSTM layers contain 1024, 512 and 256 hidden units respectively. 
The outputs of the third LSTM layer are in turn connected to an MDN. 
The LSTM layers learn to estimate the mean ($\mu$), standard deviation ($\sigma$) and weight ($\pi$) of the 3 Gaussian distributions of the MDN.
This approach has the advantage of control over the diversity and ``randomness'' of sampling, and control over the number of mixture components that allow training to account for situations where multiple predictions could be considered equally suitable. MDRNNs have previously been applied to various other tasks such as sketches~\citep{ha2017neural}, handwriting~\citep{graves2013generating}, and music control generation~\citep{Martin2019imps}.

To optimize an MDN, we minimize the negative log-likelihood of sampling true values from the predicted GMM for each example. A probability density function %
is used to obtain this likelihood value. 
This configuration corresponds to 8.5M parameters. %
The loss function in our system is calculated by the \texttt{keras-mdn-layer}~\citep{keras-mdn-layer} Python package which makes use of Tensorflow's probability distributions package to construct the PDF.
The model is trained using the Adam optimizer \citep{kingma2014adam} until the loss on the validation set failed to improve for 10 consecutive epochs.

\begin{figure*}[ht!]
     \centering
     \subfloat[The original movement sequence used to prime the model.
    \label{fig:primer}]{%
       \includegraphics[width=\textwidth,height=3cm]%
    {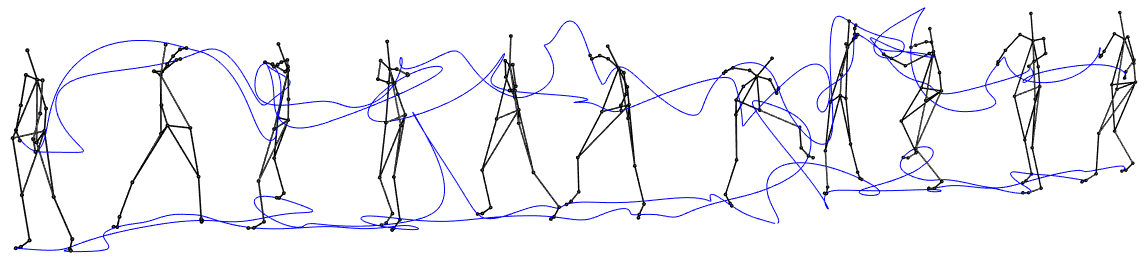}
     }
     \vfill
     \subfloat[Movement generated using the priming sequence with corresponding audio features. Movements are less smooth and expressive than the original recording, but the generated movement follows the priming example nicely. 
    \label{fig:standard}]{%
       \includegraphics[width=\textwidth]%
    {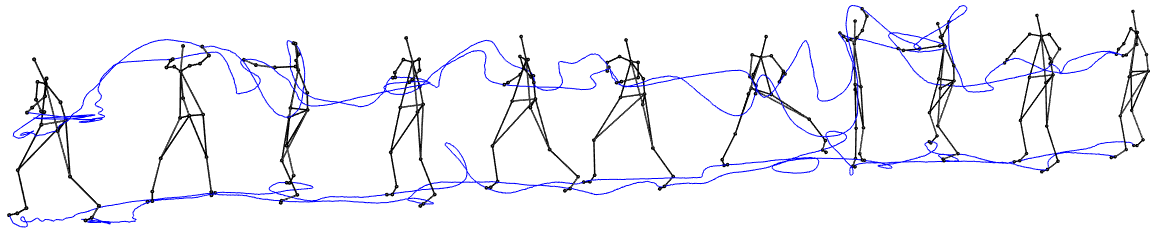}
     }
    \vfill
    \subfloat[Sequence generated when the audio features from the priming example are replaced with features from a song not used in the dataset. The movements are more unstable and shaky. 
    \label{fig:pop}]{%
       \includegraphics[width=\textwidth,height=3.2cm]%
    {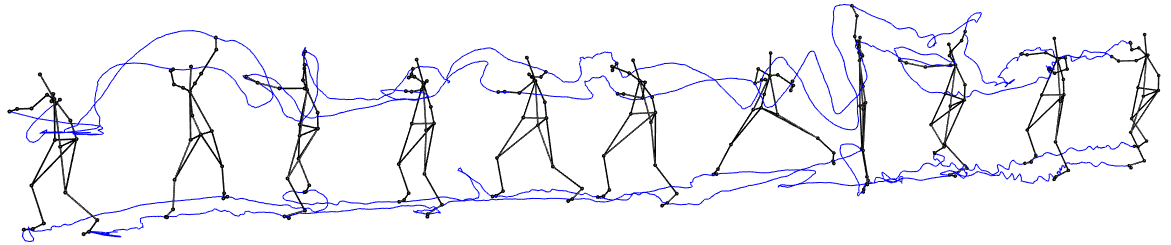}
     }
     \vfill
     \subfloat[Here, audio features are replaced with features calculated from a white noise signal. While the overall sequence is similar to the priming example, the movements contain more variation between frames, causing jittery movement.
    \label{fig:noise}]{%
       \includegraphics[width=\textwidth,height=3.3cm]%
    {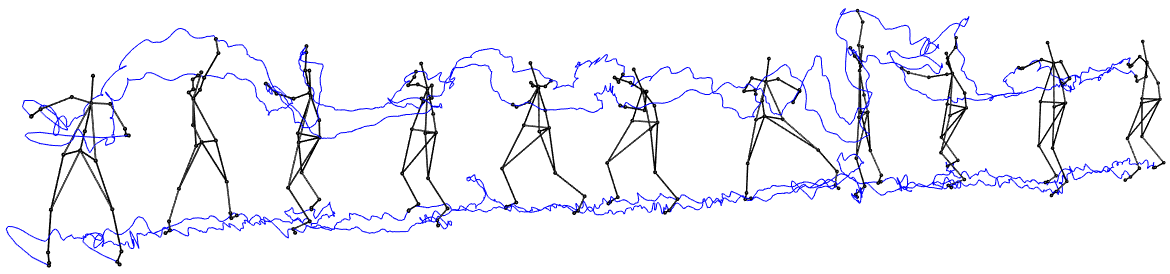}
    }
     \caption{%
     These figures show the trajectories of hand and toe markers over time (left to right). When generating motion using audio which was not used in training (\ref{fig:pop}) or white noise (\ref{fig:noise}) the movements become more unstable.}\label{fig:primimg_w_sound}
   \end{figure*}

\section{Altering Movement Using Audio Features}

When the MDRNN is trained on movement without the addition of audio features it is able to generate movements which are, under visual inspection, realistic.
In this section, we examine the effect of altering the audio input for a model trained on both movement and audio data.

To examine to what extent the model has learned a correlation between the audio features and the movement we generate motion using a priming technique. 
When using priming the input to the model is taken from one of the examples which were withheld during training.
At each time step, the input consists of the 3D positions of the 22 points and the corresponding set of audio features for that time step. 
The model then predicts the positions of the 22 points at the next time step. 
Thereby, the model always predicts the next pose using the values from the priming sequence. 
By altering the audio features of the priming example the output can be evaluated to determine the effect which different audio features have on the generated output. 
We examine three such cases here. 
First, we investigate a sequence generated using the audio features associated with the priming example itself, that is, features calculated from the musical stimuli the dancer was improvising to when the priming example was recorded.
Secondly, we replace this audio with an excerpt from a song which was not part of the training data. Finally, features calculated from a white noise signal is used to replace the original audio features.
Figure \ref{fig:primimg_w_sound} %
show keyframes from the 3 generated movement sequences as well as the motion sequence used to prime the model. %

\section{Discussion}
When generating movement using the audio features
belonging to the priming sequence (Figure \ref{fig:standard}), the model generates movement which largely follows the example (Figure \ref{fig:primer}). 
While the overall movement sequence is similar, some expressiveness seems to have been lost. 
The generated motion could be said to resemble a \say{dead pan} performance of the original sequence, as trajectories of arms and legs are to some extent muted in comparison to the original. This may be due to the model generalising movement across the training data. 

Figure \ref{fig:pop} shows the sequence generated when the original audio features are replaced with features calculated from a song not included in the training data. 
The model produces a movement sequence which matches the priming sequence well, indicating that the model is able to predict the next frame of the motion data even with unseen audio features. 
Still, additional noise is visible in this example (when compared to \ref{fig:standard}), suggesting that the model has not fully learned to generate smooth movements when unseen audio features are used. 

In the final figure, \ref{fig:noise}, the audio features are replaced with features calculated from a white noise signal.
Here, trajectories are decidedly affected by the audio. As with figure \ref{fig:pop} and \ref{fig:standard} the model still predicts reasonable positions for the 22 markers at every frame, but with a larger variation between frames, causing the resulting sequence to display jittery movement. This indicates that the model does rely on structured audio to generate realistic movements at the micro (if not macro) scale.

\section{Conclusions and Future Work}

These results indicate that the MDRNN model could be used to explore how music and audio features affect the way the dancers move, and how this manifests itself in the movements generated by a deep neural network. 
Much work remains to obtain a comprehensive understanding of how MDRNNs can model cross-modal interactions like those between sound and motion. An important aspect is how we can best evaluate the performance of this model.
Finding good qualitative and quantitative ways to evaluate creative data generated by models such as this one will be a central question in our future work. 
Going forward, we will focus on systematically exploring metrics to evaluate the generated movements, train the model on a larger dataset and experiment with alternate audio representations.

\bibliographystyle{iccc}
\bibliography{iccc2020}

\end{document}